\documentclass[12pt]{iopart}
\usepackage{color}
\usepackage{graphicx}
\usepackage{amsfonts}
\usepackage{amssymb}
\usepackage{cite}
\usepackage{iopams}
\usepackage{ulem}

\begin{document}

%----------------------------------------------------------------Title--------------------------------------------------------------------------------------------%
\title{Quantum frequency conversion of memory-compatible single photons from 606 nm to the telecom C-band}
\author{Nicolas Maring$^1$, Dario Lago-Rivera$^1$, Andreas Lenhard$^{1}$, Georg Heinze$^1$, Hugues de Riedmatten$^{1,2}$}
\address{{$^1$ ICFO-Institut de Ciencies Fotoniques, The Barcelona Institute of Science and Technology, 08860 Castelldefels (Barcelona), Spain\\
$^2$ ICREA-Instituci\'{o} Catalana de Recerca i Estudis Avan\c cats, 08015 Barcelona, Spain\\}}
\ead{nicolas.maring@icfo.eu}
\begin{abstract} 
The coherent manipulation of the frequency of single photons is an important requirement for future quantum network technologies. It allows for instance quantum systems emitting in the visible range to be connected to the telecommunication wavelengths, thus extending the communication distances. Here we report on quantum frequency conversion of memory-compatible narrow-bandwidth photons at 606 nm to the telecom C-band at 1552$\,$nm. The 200$\,$ns long photons, compatible with Praseodymium-based solid-state quantum memories are frequency converted using a single-step difference frequency generation process in a periodically poled Lithium Niobate waveguide. We characterize the noise processes involved in the conversion and by applying strong spectral filtering of the noise, we demonstrate high signal-to-noise ratio conversion at the single photon level (SNR$\,>\,$100 for a mean input photon number per pulse of 1). We finally observe that a memory compatible heralded single photon with a bandwidth of 1.8$\,$MHz, obtained from a spontaneous parametric down conversion pair source, still shows a strong non-classical behavior after conversion. We first demonstrate that correlations between heralding and converted heralded photons stay in the non-classical regime. Moreover, we measured the heralded autocorrelation function of the heralded photon using the converter device as a frequency-domain beam splitter, yielding a value of 0.19$\pm0.07$. The presented work represents a step towards the connection of several quantum memory systems emitting narrow-band visible photons to the telecommunication wavelengths.
\end{abstract}

%\pacs{}

%-------------------------------------------------------------Body-----------------------------------------------------------------------------------------------%

\section{Introduction}
The ability to control the optical frequency of quantum state carriers (i.e. photons) is an important functionality for future quantum networks \cite{Kimble2008}. It allows all matter quantum systems -- nodes of the network -- to be compatible with the telecommunication C-band, therefore enabling long distance fiber quantum communication between them. It also allows dissimilar nodes to be connected with each other, thus resulting in heterogeneous networks that can take advantage of the different capabilities offered by the diversity of its constituents \cite{Walmsley2016}. 
Quantum frequency conversion (QFC) \cite{Kumar1990} aims for the energy shift of an optical field, in a coherent and noise-free fashion, such that quantum properties are preserved to a high degree. Optical frequency conversion based on three or four-wave mixing, has been demonstrated using different platforms such as non-linear waveguides \cite{Langrock2005,Tanzilli2005}, non-linear crystals in cavities \cite{Albota2004,Samblowski2014}, microresonators \cite{Li2015,Guo2016} or atomic systems \cite{Radnaev2010,Bustard2017}. The high conversion efficiencies offered by these platforms, led to multiple experiments showing the increase of telecom detection efficiencies \cite{Albota2004}, generation of non-classical states of light at a specific wavelength \cite{Vollmer2014,Kong2014}, connections of quantum systems to the telecommunication band for long distance communication application \cite{Takesue2010,Curtz2010,Albrecht2014,Radnaev2010,Zaske2012,Maring2014,Farrera2016a,Ikuta2016,Bock2017,Ikuta2017}, or the coupling of disparate quantum memory (QM) systems \cite{Maring2017}. The main challenge of QFC is to achieve a high conversion efficiency together with a low noise generation at the target wavelength, given that the required pump light often generates a high amount of uncorrelated noise photons. 
\\
Depending on the wavelengths of the input signal and the converted photons, we can distinguish between three different cases of noise generation at the target wavelength in non-linear materials \cite{Pelc2011a}. One in which the pump field frequency is far below the frequency of signal and converted photons, thus making noise-free frequency conversion possible \cite{Kong2014,Ates2012,Lenhard2017}. A second case in which one of the two photons frequencies are close to the pump, where Raman noise is generated \cite{Fernandez-Gonzalvo2013,Zaske2012}. A third case in which the pump field is in between the signal and converted photons and also generates non-phase matched parametric fluorescence noise \cite{Pelc2010} on top of Raman noise. Note that cascaded conversion using a long wavelength pump, as demonstrated in Ref.~\cite{Pelc2012} for up-conversion, could be a solution to avoid noise generation at the target wavelength.
\\
%In the context of quantum networks several systems have been proposed combining source of light-matter entanglement and a QM in a single physical system. These are known as emissive quantum nodes \cite{Afzelius2015}.
In the context of quantum networks several systems, known as emissive quantum nodes, combine the generation of light-matter entanglement with quantum storage capabilities \cite{Afzelius2015}.
An important challenge is therefore the conversion of narrowband light emitted by these nodes to telecom wavelengths. The conversion of such narrow-band photons indeed leads to stringent requirements for noise reduction, as temporal gating cannot be used due to their long temporal waveform. Several important quantum node systems emit light around 600 nm, e.g. Europium doped QMs (580$\,$nm) \cite{Laplane2017}, Praseodymium doped QMs (606$\,$nm) \cite{Kutluer2017}, and NV centers (637$\,$nm) \cite{Hensen2015}. Single-step conversion requires a pump wavelength close to 1$\,\mu m$, which generates non-phase matched noise photons at the target-wavelength. Several works have demonstrated QFC from telecom to the visible of broadband photons \cite{Albota2004,Rutz2017,Allgaier2016}. In Ref. \cite{Maring2014}, we demonstrated QFC from 1570 to 606$\,$nm of QM-compatible weak coherent states at the single photon level, using the QM as ultra-narrowband filter. First attempts have been made towards the QFC of NV center resonant light to telecom \cite{Pelc2010,Ikuta2014}, however without demonstrating quantum light conversion.

In this paper, we demonstrate efficient direct frequency conversion of memory-compatible photons at 606 nm to the telecom C-band based on difference frequency generation (DFG) in a non-linear waveguide. We characterize the noise processes in the device, limiting the single photon level operation of the frequency conversion, and apply strong filtering to obtain high signal-to-noise ratio (SNR) for weak coherent pulses compatible with a Pr$^{3+}$:Y$_2$SiO$_5$ doped quantum memory. Finally we employ an ultra-narrow band photon pair source to demonstrate quantum frequency conversion of a memory-compatible heralded single photon.
To that end, we measure the cross-correlation between heralding and converted heralded photons. We also measure the heralded photon autocorrelation in a Hanbury-Brown-Twiss configuration using the converter device as a frequency-domain beam splitter.

\section{The frequency conversion device}
\subsection{Concept}
The concept of QFC can be described considering two optical modes s and c (signal and converted), at frequencies $\omega_s$ and $\omega_c$ respectively. In the case of difference frequency generation, it satisfies $\omega_c = \omega_s - \omega_P$, where $\omega_P$ is the frequency of the pump field. In the limit of an undepleted pump field, the Hamiltonian describing the three wave mixing process is \cite{Kumar1990,Ikuta2011}

\begin{equation}\label{eqn:hamiltonian}
\hat{\mathcal{H}} = i\hbar \left(\chi^* \hat{a}^\dagger_c \hat{a}_s - \chi \hat{a}_c \hat{a}^\dagger_s \right),
\end{equation} 
where $\hat{a}_c$ ($\hat{a}_s$) is the annihilation operator of mode c (s), and $\chi$ is proportional to the pump field amplitude. This Hamiltonian can be interpreted as a frequency domain beam splitter \cite{Ikuta2011} where the transmittance and reflection of the modes s and c depend on the pump amplitude applied.

\subsection{Setup}

\begin{figure}
	\centering\includegraphics[width=.65\textwidth]{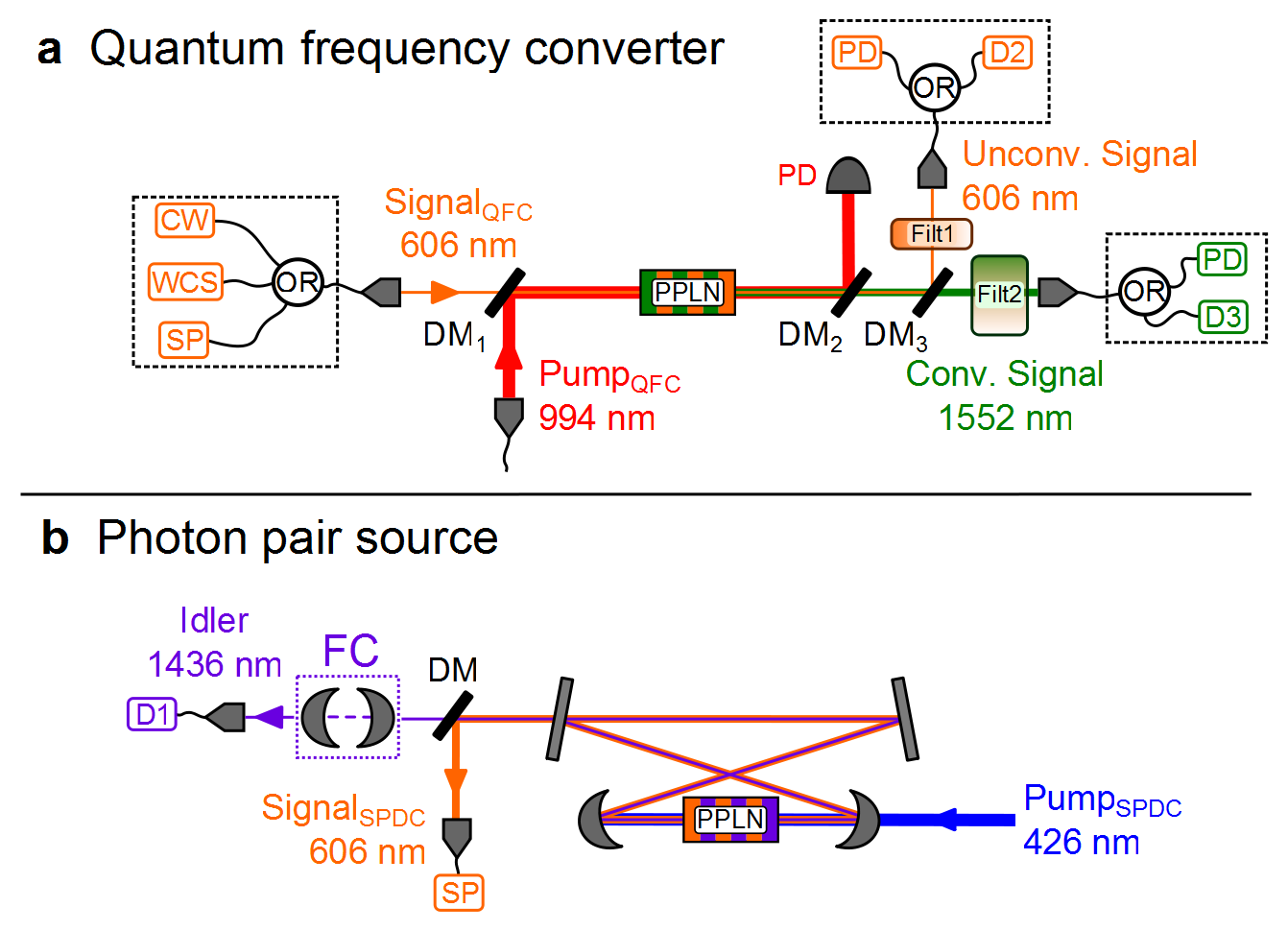}
	\caption{(a) QFC setup. At the input of the device, either Continuous Wave (CW), Weak Coherent States (WCS) or heralded Single Photons (SP) at 606 nm can be coupled together with a strong 994 nm pump into the Periodically Poled Lithium Niobate (PPLN) waveguide. At the output, the pump, the unconverted and the converted lights are separated by means of Dicroic Mirrors (DM). The coupled pump is monitored with a Photodiode (PD). The unconverted and the converted fields go through filtering stages and are coupled into a single mode fiber. Finally they are either monitored with photodiodes, or detected with single-photons detectors (D2 and D3) (b) Source setup. The photon pair source consists of a PPLN crystal inside a bow-tie cavity. It is pumped with a 426 nm continuous-wave laser beam. The generated idler photon at 1436 nm and signal photon at 606 nm are doubly resonant with the cavity. At the output of the cavity they are separated by means of a dicroir mirror. The idler photon passes through a Filter Cavity (FC), is then coupled to a single mode fiber and finally detected at D1 (Id220 ID Quantique, 10$\,\%$ detection efficiency, 400$\,$Hz dark counts). The 606 photon is also coupled to a single mode fiber and sent to the quantum frequency converter.}
	\label{Figure1}
\end{figure}

In our experiment the QFC process takes place in a 1.4$\,$cm long Periodicaly Poled Lithium Niobate (PPLN) ridge-waveguide (HC Photonics), temperature-stabilised at 65 degrees Celsius. Both facets are anti-reflection coated for the three wavelengths involved in the experiment. The setup is depicted in Fig.~\ref{Figure1}(a). A pump laser at 994 nm is coupled with 55$\,\%$ efficiency, together with a 606 nm signal with 57$\,\%$ coupling efficiency inside the waveguide. At the output of the waveguide, the converted 1552 nm light field, the 994 nm pump and finally the non-converted 606 nm signal are separated by means of dicroic mirrors and can be monitored independently. The converted signal passes through a filtering stage and is coupled into a single mode fiber (79$\,\%$ coupling efficiency). The filtering stage is composed of a fiber bragg grating (65$\,\%$ transmission, 2.5 GHz bandwidth) and an etalon (95$\,\%$ transmission, 210$\,$MHz bandwidth and free spectral range (FSR) of 4 GHz). The unconverted signal is also filtered by means of a diffraction grating (75$\,\%$ reflection) and an Etalon (90$\,\%$ transmission, 10$\,$GHz linewidth, FSR of 60$\,$GHz) and also coupled into a single mode fiber. At the input of the device, we can choose between different 606 nm light fields: bright laser light, weak coherent states (WCS) at low average photon number per pulses, or heralded single photons generated by a photon-pair source.

\subsection{QFC performance}
\begin{figure}
	\centering\includegraphics[width=0.65\textwidth]{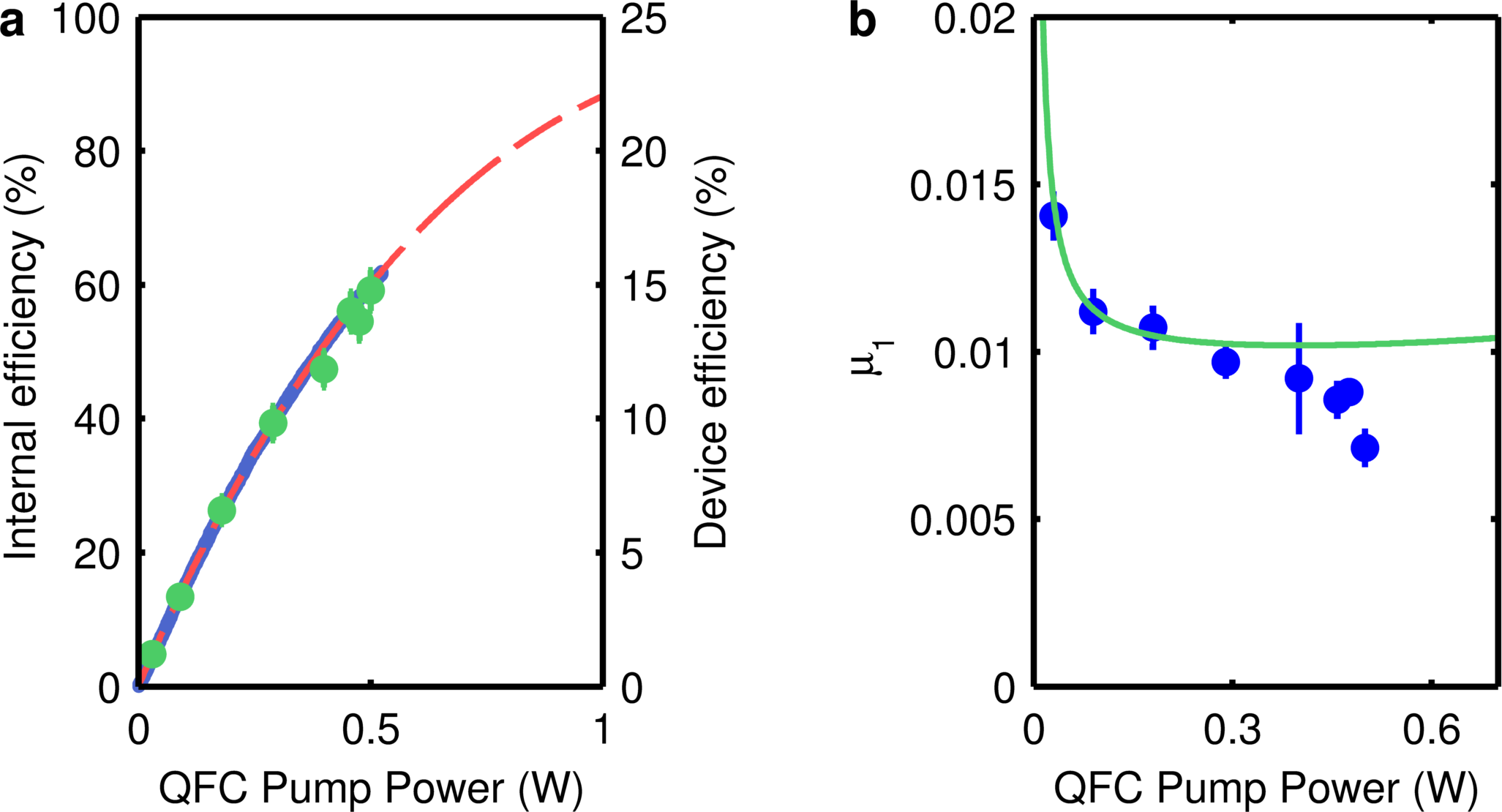}
	\caption{QFC performance. (a) Conversion efficiency from 606$\,$nm to 1552$\,$nm as a function of the coupled pump power. The blue curve shows the efficiency using bright classical light and sweeping the coupled pump power from 0 to 530$\,$mW in 100$\,\mu s$. The red dashed curve shows the fit of the efficiency using eq(2). The greens points show the efficiency measured with weak coherent states. (b) $\mu_1$ measurements of the converted weak coherent states as a function of the coupled pump power. The green curve shows the expected values of the $\mu_1$.}
	\label{Figure2}
\end{figure}

The frequency converter is first characterized using classical light as input. The conversion efficiency of the device is measured as a function of the coupled pump power inside the waveguide, using 1 mW of classical input light at 606 nm. For this measurement only, the pump power is swept using an Acousto Optical Modulator (AOM) over a short time of 100 $\mu s$. The coupled pump power and the difference-frequency converted light are then monitored at the output of the waveguide using photodiodes. This fast measurement offers the advantage of avoiding thermal effects (change of the quasi-phase matching temperature due to local heating of the waveguide at high pump powers) sometimes encountered when changing pump power \cite{Rutz2016a}, and gives a more precise efficiency measurement. The blue trace in Fig.~\ref{Figure2}(a) shows the measured internal efficiency depending on the coupled pump power, inferred by correcting for all losses. At the maximum coupled pump power available of 530 mW, we measure an internal conversion efficiency of 62 $\%$. This value is in accordance with the measured depletion of the 606 nm signal. The device efficiency, also shown on the right axis, includes all optical losses: signal transmission (93$\,\%$), coupling efficiency of the signal in the waveguide (57$\,\%$), filtering efficiency  (62$\,\%$) and fiber coupling efficiency (79$\,\%$) of the converted signal. The conversion efficiency is fitted with the model \cite{Roussev2004}

\begin{equation}
\eta_{QFC} = \eta_{max} \sin^2\left(L\sqrt{\eta_nP_c}\right),
\end{equation} 
where $\eta_{max}$ is the maximum efficiency, \textit{L} is the length of the waveguide, $\eta_n$ is the normalized efficiency and $P_c$ is the coupled pump power. The fit gives a maximum internal efficiency of $\left(95\pm0.1\right)\,\%$ (24.5 $\%$ device efficiency) at 1.45$\,$W of coupled pump power and a normalized efficiency of $\left(86.1 \pm 0.1\right)\,\%$W$^{-1}$cm$^{-2}$.

To assess the potential of our QFC device for converting light emitted by quantum memories, we also characterize it at the single-photon level, sending 606 nm weak coherent states with 200$\,$ns FWHM Gaussian shape, mimicking long single photons compatible with Praseodymium doped quantum memories \cite{Gundogan2015,Rielander2014,Seri2017}. For this measurement and the following ones, the 994$\,$nm pump runs in continuous and the waveguide temperature is optimized for each pump power. The converted photons are then detected with an InGaAs single photon detector D3 (ID230, ID Quantique, 10$\,\%$ detection efficiency, 10$\,$Hz dark counts) and integrated over a 400$\,$ns time window, containing more than 98$\,\%$ of the pulses. The device efficiency is also extracted from this measurement and plotted in Fig.~\ref{Figure2}(a), matching with the classical measurement. We also measured the signal to noise ratio (SNR) of the converted photons with different average input photon numbers per pulse, ranging from 0.04 to 1. From a linear fit of this measurement we extract the parameter $\mu_1$, i.e. the number of photons per pulse at the input of the device in order to obtain a SNR of 1 at the output \cite{Fernandez-Gonzalvo2013}. Fig.~\ref{Figure2}(b) shows the measurement of $\mu_1$ as a function of the pump power. Interestingly, we observe a decrease in $\mu_1$ with increasing pump power, down to $\left(7\pm0.5\right)\times10^{-3}$ at 500$\,$mW pump power. Although not intuitive, the decrease of $\mu_1$ (i.e increase of SNR) with the pump power can be explained by characterizing the noise processes in the waveguide, as discussed in the next section. The green curve shows the expected values of the $\mu_1$, calculated from the classical conversion efficiency measurement, and the noise level (Fig.~\ref{Figure3}(b)). This measurement demonstrate the capability of our device to convert long photons with high signal to noise ratio, thanks to the high efficiency of the process and to the strong spectral filtering of the QFC noise.

\subsection{Noise Processes}

\begin{figure}
	\centering\includegraphics[width=0.65\textwidth]{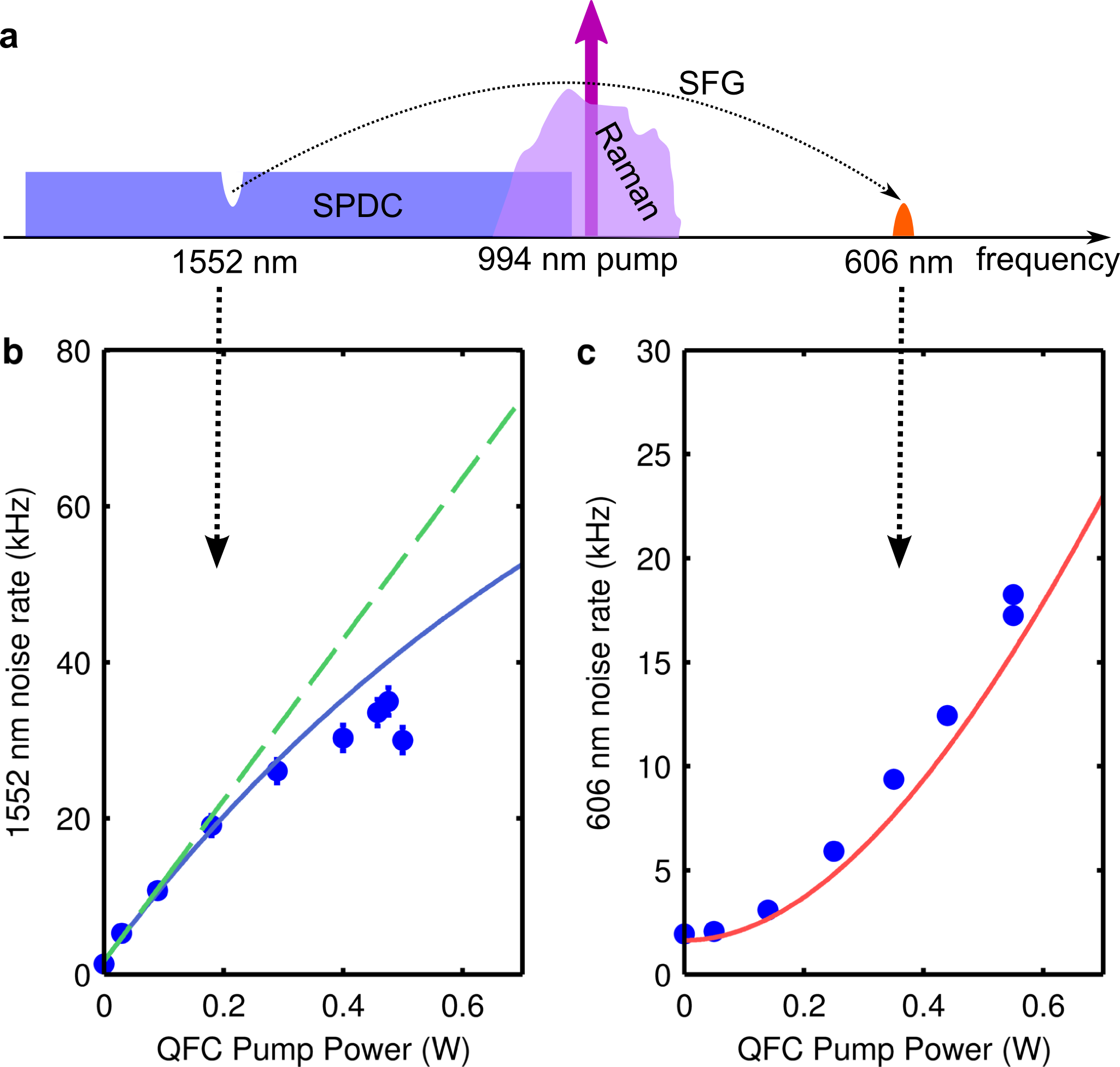}
	\caption{QFC noise characteristic. (a) Schematic of the different noise processes induced by a strong pump field in the waveguide. (b) Measurement of noise in the 1552 nm region, dominated by Spontaneous Parametric Down Conversion (SPDC) noise photons, as a function of the pump power coupled in the waveguide. (c) Noise measurement in the 606 nm region. The two measurements are performed gating the single-photon detector and using the narrow band filters shown in the setup section (of 10$\,$GHz bandwidth for the 606 nm photons and 210$\,$MHz bandwidth for the converted 1552 nm photons). The data are normalized as counts per second at the output of the waveguide over 1$\,$GHz bandwidth assuming a continuous measurement (i.e. eliminating temporal gate).}
	\label{Figure3}
\end{figure}
We now study in more detail the noise generation inside the converter. The noise processes are depicted in Fig.~\ref{Figure3}(a). A strong pump field generally generates Raman noise around its central frequency. In a PPLN waveguide it has been estimated to have a width smaller than 30$\,$THz  (1000$\,$cm$^{\mathrm{-1}}$) \cite{Pelc2011a,Zaske2011}. Raman noise therefore does not play a significant role in our experiment, since the frequency shift between pump and signal is 109$\,$THz. A second type of noise is Spontaneous Parametric Down Conversion (SPDC) noise as described in Ref. \cite{Pelc2010}, generated at lower frequencies than the pump. In our case we observe direct SPDC noise at the converted wavelength of 1552 nm ($\lambda>\lambda_{pump}$). Part of this noise, within the phase matching bandwidth of the frequency converter, can be eventually converted to 606$\,$nm by the pump field via sum frequency generation. In the case of $\lambda<\lambda_{pump}$ the expected behavior is therefore very different as the reconversion of the noise leads to a quadratic dependence with pump power.
We measure the noise using only the pump as the input of the waveguide, and detecting the photons either at the 1552$\,$nm output (Fig.~\ref{Figure3}(b)) or at the 606$\,$nm output (Fig.~\ref{Figure3}(c)). The telecom noise saturates as a function of the pump power due to the above explained back-conversion to 606$\,$nm. This gives approximately a factor 2 reduction of the expected noise at the maximum pump power and explains why the SNR is not decreasing with the pump power. Taking the first 3 points to fit a linear slope, we find an internal SPDC noise generation coefficient of $\alpha_N$=76$\,$kHz/mW/cm normalized to a 1$\,$THz bandwidth, similar as the one described in Ref. \cite{Ikuta2014}. The blue curve, matching our data, shows a model that takes into account the generation of noise $\alpha_N$ along the waveguide at the position $x$ and its eventual back-conversion on the remaining length of the waveguide.

\begin{equation}
\label{eqnoise}
N_{telecom}(P) = \alpha_NP \int_0^L \left(1-\eta_{max} \sin^2\left((L-x)\sqrt{\eta_nP}\right)\right)dx,
\end{equation}

where the parameters $\eta_{max}$ and $\eta_n$ are the efficiencies found with the classical conversion efficiency measurement. 
The converted noise detected at 606$\,$nm shows a clear quadratic behavior, as expected. The red line shows the expected back-converted noise level using the same model as before, i.e. eqn.~(\ref{eqnoise}).

\section{Conversion of heralded single photons}
\begin{figure}
	\centering\includegraphics[width=0.65\textwidth]{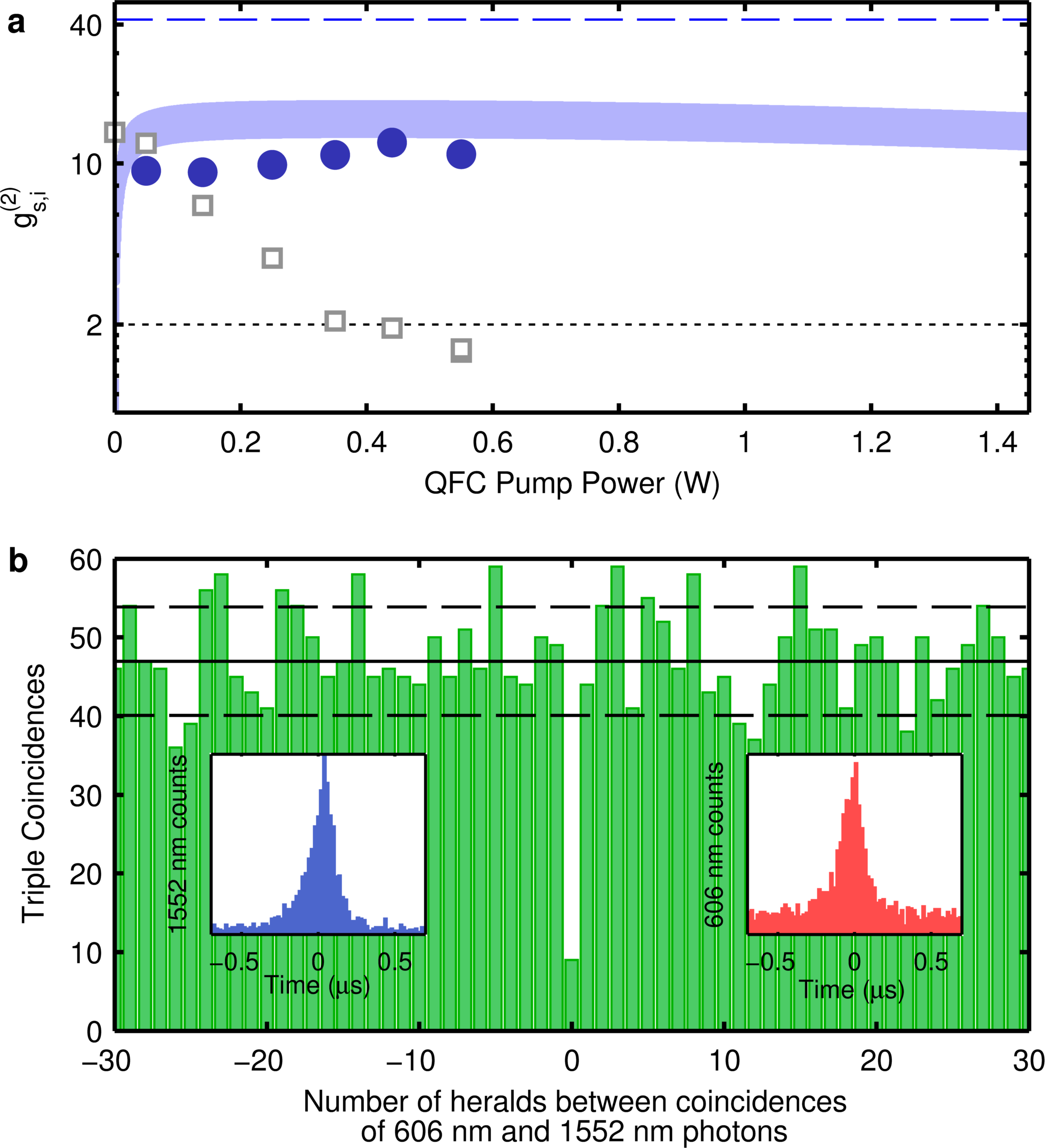}
	\caption{Non-classical correlations. (a) Cross correlation measurement between the 1436 nm heralding photon and the converted 1552 nm photon (blue dots) The blue shaded area shows the expected values taking into account the SNR of the converted light and the estimated cross-correlations for the source in single mode before the conversion. The grey open squares show the depleted 606$\,$nm photon as a function of the coupled pump power of the QFC. Error bars are smaller than the points. The dotted line represents the classical threshold of 2. The dashed blue line represents the measured cross-correlation of the source in single mode configuration for signal and idler. (b) Histogram of the triple coincidence between the heralding photon, the 1552 converted photon and the 606 unconverted photon, measured with 250 mW of coupled QFC pump power. The left (right) inset shows the histogram of the heralded converted (unconverted) photons.}
	\label{Figure4}
\end{figure}
To prove that our QFC device can operate with memory compatible quantum light, we use a photon pair source, schematically depicted in Fig.~\ref{Figure1}(b).
\subsection{The photon pair source}
The source is a new generation of an earlier source \cite {Fekete2013, Rielander2016} that has been used to to demonstrate quantum storage of heralded single photons in a Pr$^{3+}$:Y$_2$SiO$_5$ memory \cite{Rielander2014, Seri2017}. It is based on a 2$\,$cm long PPLN crystal, placed in a bow-tie cavity (FSR of 261$\,$MHz). Pumped with 426$\,$nm light in CW, it generates a signal photon at 606$\,$nm, and an idler photon at 1436$\,$nm. The biphoton linewidth is 1.8$\,$MHz, making the 606$\,$nm photon compatible for storage in a Pr$^{3+}$:Y$_2$SiO$_5$ based quantum memory \cite{Rielander2014,Seri2017}. The idler telecom photon is filtered with a Fabry-Perot cavity (linewidth of 80$\,$MHz, FSR = 17$\,$GHz) in order to select a single frequency mode out of the 8 modes at the output of the bow-tie cavity. It is then used to herald the signal 606$\,$nm photon with an efficiency $\eta_{\mathrm{S}}^{\mathrm{h}}$ of 25$\,\% $ in single mode fiber. In the 606$\,$nm arm, there is no filtering of a single frequency mode and the non-correlated modes contribute to accidental coincidences. Pumped with 1.65$\,$mW of 426$\,$nm light, the source generates about 280 heralded 606$\,$nm photons per second which are strongly non-classically correlated to the heralding photon. This number is limited by the transmission of the heralding photons (filtering cavity and fiber coupling) and their detection at D1 (10$\,\%$ efficiency). The correlation time of the photon pair is measured to be 120.9$\,$ns. We measure the normalized cross correlation function $g_{s,i}^{(2)} = \frac{P_{s,i}}{P_sP_i}$,where $P_{s,i}$ describes the probability for a coincidence detection of a signal and an idler photon, and $P_s$ ($P_i$) is the detection probabilities for single signal (idler) events. Using a detection window of 400$\,$ns we obtain $g_{s,i}^{(2)} = $15.9$\pm$0.9, well above the classical threshold of 2, assuming thermal statistics for the signal and idler fields. The single photon nature of the heralded 606$\,$nm photon is verified measuring the heralded second-order autocorrelation function $g^{(2)}_{H,source}$. This is done using a 50/50 fiber-beam splitter in the 606$\,$nm arm and detecting the heralded correlations between the two outputs. The source exhibits $g^{(2)}_{H,source}$ = 0.12 $\pm 0.01$, well below the classical threshold of 1, and below the threshold of 0.5 for a two photon Fock state.

\subsection{Results}
Finally, we connect the photon pair source to the quantum frequency converter. The correlations between the herald and the non-converted signal (detectors D1 and D2), and between the herald and the converted signal (detectors D1 and D3) are measured as a function of the QFC pump power and are shown in Fig.~\ref{Figure4}(a). At 0$\,$mW pump power, the normalized cross correlation function of 13.6$\pm0.9$ for the non-converted light (grey open squares) corresponds to the source without any effect of the QFC (except for the additional transmission losses). The $g^{(2)}_{s,i}$ value for the non-converted photons then rapidly drops when increasing the pump power due to the drop of SNR induced by the high amount of quadratic noise generated through the 10$\,$GHz etalon filter.
In contrast, when looking at the 1552$\,$nm converted signal (blue dots) we observe that the value of $g^{(2)}_{s,i}$ slightly increases with pump power up to 12.3 $\pm 0.7$ at 440 mW. It is worth mentioning that the filtering stage of the converted signal used to obtain a high SNR of the converted photon also filters a single frequency mode from the bow-tie cavity. It then reduces the number of accidentals coincidences from the source, as the other modes, non-correlated with the single frequency idler mode, are not detected. On the contrary, the QFC itself adds noise to the channel, thus reducing the cross correlation function. Thus it is difficult to directly compare the converted cross correlation function with the non-converted one shown in the Figure~\ref{Figure4}(a). Instead we measured the cross-correlation $g^{(2)}_{s,i}$ of the source before the QFC sending the 606$\,$nm heralded single-photon through a 12$\,$MHz transparency window of a Pr$^{3+}$:Y$_2$SiO$_2$ doped crystal \cite{Seri2017}, that filters a single frequency mode of the bow tie cavity and measure $g^{(2)}_{s,i} = $ 42$\pm$7. We now have an estimate of the performance of the source in fully single mode operation that can be compared with the converted one measured (blue dots in Fig.~\ref{Figure4}(a)). The effect of the noise of the QFC device on correlations can be estimated using the same approach as in Ref.\cite{Albrecht2014}: 
\begin{equation}
g^{(2)}_{c,i} = g^{(2)}_{s,i} \frac{\eta_{\mathrm{S}}^{h}/\mu_1+1}{\eta_{\mathrm{S}}^{h}/\mu_1+g^{(2)}_{s,i}},
\end{equation}
where $g^{(2)}_{c,i}$ ($g^{(2)}_{s,i}$) is the cross-correlation function of the converted (unconverted) photon. We plot the expected correlations after conversion (Blue shaded area in Fig.~\ref{Figure4}(a)) using the $\mu_1$ model shown previously (green curve of Fig.~\ref{Figure2}(b)). We can then observe that, up to the pump power of 1.45$\,$W for which the maximum conversion efficiency should occur, the correlations, although degraded by the QFC-induced noise, remain well above the classical limit of 2. The model suggests that the converted light can exhibit non-classical correlations at any pump power. The small discrepancy between the measured data and the model is probably due to an overestimation of the non-converted correlations as the filtering is much narrower in that case.

In order to show the preservation of the single photon nature of the converted heralded photon, we measure -- for the first time to our knowledge -- the heralded-autocorrelation function of the signal photon using the QFC as a frequency-domain beam splitter \cite{Clemmen2016,Kobayashi2016}. The pump power of the QFC could in principle be tuned in such a way that the photon entering the waveguide has the same probability of being converted or staying in its original state (eq.~\ref{eqn:hamiltonian}). 
In our case, we equalize the photon detection rates between the two outputs of the frequency beam splitter by fixing the QFC pump power at 250$\,$mW, leading to a conversion efficiency of 35$\,\%$. This also dramatically reduces the amount of quadratic noise detected at D2. We can then record the triple coincidences between the heralding photon at D1, the unconverted 606 photon at D2 and the converted 1552 nm photon at D3 to measure the heralded autocorrelation function $g^{(2)}_{s,c|i}$ in a 400 ns window. The histogram of the triple coincidence is shown in Fig.~\ref{Figure4}(b), sorted by the number of heralding events between succeeding detections at signal or converted photon detectors \cite{Fasel2004}. The value at bin 0 corresponds to $g^{(2)}_{s,c|i}$ of 0.19$\pm0.07$. It is well below the classical threshold and proves the single photon nature of the converted light. Note that this value is an upper bound as a high amount of uncorrelated noise is added to the non-converted output of the beam splitter. 
This measurement also highlights the potential of a quantum frequency converter as a beam splitter device, which for instance could be used to perform Bell measurements between modes of different wavelength \cite{Raymer2010}.

\section{Conclusion}
In conclusion, we developed a memory compatible quantum frequency conversion device bridging the gap between visible light around 600 nm and the telecom C-band. The device shows high intrinsic conversion efficiency and low noise at the target wavelength. The device conversion efficiency could be greatly improved using a longer waveguide as well as more efficient filtering or coupling techniques. Note that despite a measured device efficiency of 15$\,\%$, this conversion process becomes advantageous after only 1$\,$km of fiber transmission of 606$\,$nm photons. We showed conversion of quantum memory-compatible photons, at the single photon level, with high signal to noise ratio. We finally demonstrated that single photons compatible with a Pr$^{3+}$:Y$_2$SiO$_5$ quantum memory still exhibit strong quantum correlations after the conversion.The degradation of quantum properties is only due to the noise generated by the QFC. This issue could be tackled with stronger filtering. Our work opens the route towards connecting different solid-state quantum memory systems emitting in the visible range (e.g. Europium and Praseodymium quantum memories or NV centers) to the telecom C-band.

Note added: We became aware of a related publication demonstrating quantum frequency conversion from NV centers \cite{Dreau2018}.

\section*{Funding Information}
This work was supported by the ERC through the starting grant QuLIMA, by the Spanish Ministry of Economy and Competitiveness (MINECO) and the Fondo Europeo de Desarrollo Regional (FEDER) through grant FIS2015-69535-R, by MINECO Severo Ochoa through grant SEV-2015-0522 by Fundaci\'{o} Privada Cellex and by the CERCA programme of the Generalitat de Catalunya. G.H. acknowledges support by the ICFOnest international postdoctoral fellowship program.
\section*{Acknowledgements}
We thank Margherita Mazzera and Daniel Riel{\"{a}}nder for helpful discussions and for their contributions at the early stage of the experiment.

%----------------------------------------------------------------References--------------------------------------------------------------------------------------------%
\section*{References}
\bibliographystyle{osajnl}

\begin{thebibliography}{10}
\newcommand{\enquote}[1]{``#1''}	
	
\bibitem{Kimble2008}
H.~J. Kimble, \enquote{{The quantum internet},} Nature \textbf{453}, 1023--1030
(2008).

\bibitem{Walmsley2016}
I.~A. Walmsley and J.~Nunn, \enquote{{Editorial: Building Quantum Networks},}
Physical Review Applied \textbf{6}, 040001 (2016).

\bibitem{Kumar1990}
P.~Kumar, \enquote{{Quantum frequency conversion},} Optics Letters \textbf{15},
1476 (1990).

\bibitem{Langrock2005}
C.~Langrock, E.~Diamanti, R.~V. Roussev, Y.~Yamamoto, M.~M. Fejer, and
H.~Takesue, \enquote{{Highly efficient single-photon detection at
		communication wavelengths by use of upconversion in reverse-proton-exchanged
		periodically poled LiNbO3 waveguides},} Optics Letters \textbf{30}, 1725
(2005).

\bibitem{Tanzilli2005}
S.~Tanzilli, W.~Tittel, M.~Halder, O.~Alibart, P.~Baldi, N.~Gisin, H.~Zbinden, \enquote{{A photonic quantum information interface},} Nature \textbf{437}, 116--120 (2005).

\bibitem{Albota2004}
M.~A. Albota and F.~N.~C. Wong, \enquote{{Efficient single-photon counting at
		155 µm by means of frequency upconversion},} Optics Letters \textbf{29},
1449 (2004).

\bibitem{Samblowski2014}
A.~Samblowski, C.~E. Vollmer, C.~Baune, J.~Fiur{\'{a}}{\v{s}}ek, and
R.~Schnabel, \enquote{{Weak-signal conversion from 1550 to 532 nm with 84{\%}
		efficiency},} Optics Letters \textbf{39}, 2979--2981 (2014).

\bibitem{Li2015}
Q.~Li, M.~Davan{\c{c}}o, and K.~Srinivasan, \enquote{{Efficient and low-noise
		single-photon-level frequency conversion interfaces using silicon
		nanophotonics},} Nature Photonics \textbf{10}, 406--414 (2016).

\bibitem{Guo2016}
X.~Guo, C.-L. Zou, H.~Jung, and H.~X. Tang, \enquote{{On-Chip Strong Coupling
		and Efficient Frequency Conversion between Telecom and Visible Optical
		Modes},} Physical Review Letters \textbf{117}, 123902 (2016).

\bibitem{Radnaev2010}
A.~G. Radnaev, Y.~O. Dudin, R.~Zhao, H.~H. Jen, S.~D. Jenkins, A.~Kuzmich, and
T.~a.~B. Kennedy, \enquote{{A quantum memory with telecom-wavelength
		conversion},} Nature Physics \textbf{6}, 894--899 (2010).

\bibitem{Bustard2017}
P.~J. Bustard, D.~G. England, K.~Heshami, C.~Kupchak, and B.~J. Sussman,
\enquote{{Quantum frequency conversion with ultra-broadband tuning in a Raman
		memory},} Physical Review A \textbf{95}, 053816 (2017).

\bibitem{Vollmer2014}
C.~E. Vollmer, C.~Baune, A.~Samblowski, T.~Eberle, V.~H{\"{a}}ndchen,
J.~Fiur{\'{a}}{\v{s}}ek, and R.~Schnabel, \enquote{{Quantum up-conversion of
		squeezed vacuum states from 1550 to 532 nm},} Physical Review Letters
\textbf{112}, 2--6 (2014).

\bibitem{Kong2014}
D.~Kong, Z.~Li, S.~Wang, X.~Wang, and Y.~Li, \enquote{{Quantum frequency down-conversion of bright amplitude-squeezed states},} Optics Express \textbf{22}, 24192 (2014).

\bibitem{Takesue2010}
H.~Takesue, \enquote{{Single-photon frequency down-conversion experiment},} Physical Review A \textbf{82}, 013833 (2010).

\bibitem{Curtz2010}
N.~Curtz, R.~Thew, C.~Simon, N.~Gisin, and H.~Zbinden, \enquote{{Coherent frequency-down-conversion interface for quantum repeaters},} Optics Express
\textbf{18}, 22099 (2010).

\bibitem{Albrecht2014}
B.~Albrecht, P.~Farrera, X.~Fernandez-Gonzalvo, M.~Cristiani, and
H.~de~Riedmatten, \enquote{{A Waveguide Frequency Converter Connecting Rubidium Based Quantum Memories to the Telecom C-Band},} Nature Communications \textbf{5}, 1--6 (2014).

\bibitem{Zaske2012}
S.~Zaske, A.~Lenhard, C.~A. Ke{\ss}ler, J.~Kettler, C.~Hepp, C.~Arend,
R.~Albrecht, W.~M. Schulz, M.~Jetter, P.~Michler, and C.~Becher,
\enquote{{Visible-to-telecom quantum frequency conversion of light from a single quantum emitter},} Physical Review Letters \textbf{109}, 1--5 (2012).

\bibitem{Maring2014}
N.~Maring, K.~Kutluer, J.~Cohen, M.~Cristiani, M.~Mazzera, P.~M. Ledingham, and H.~de~Riedmatten, 
\enquote{{Storage of up-converted telecom photons in a doped crystal},} New Journal of Physics \textbf{16} (2014).

\bibitem{Farrera2016a}
P.~Farrera, N.~Maring, B.~Albrecht, G.~Heinze, and H.~de~Riedmatten,
\enquote{{Nonclassical correlations between a C-band telecom photon and a
		stored spin-wave},} Optica \textbf{3}, 1019 (2016).

\bibitem{Ikuta2016}
R.~Ikuta, T.~Kobayashi, K.~Matsuki, S.~Miki, T.~Yamashita, H.~Terai,
T.~Yamamoto, M.~Koashi, T.~Mukai, and N.~Imoto, \enquote{{Heralded single
		excitation of atomic ensemble via solid-state-based telecom photon
		detection},} Optica \textbf{3}, 1279 (2016).

\bibitem{Bock2017}
M.~Bock, P.~Eich, S.~Kucera, M.~Kreis, A.~Lenhard, C.~Becher, and J.~Eschner,
\enquote{{High-fidelity entanglement between a trapped ion and a telecom
		photon via quantum frequency conversion},} arXiv 1710.04866  (2017).

\bibitem{Ikuta2017}
R.~Ikuta, T.~Kobayashi, T.~Kawakami, S.~Miki, M.~Yabuno, T.~Yamashita,
H.~Terai, M.~Koashi, T.~Mukai, T.~Yamamoto, and N.~Imoto,
\enquote{{Polarization insensitive frequency conversion for an atom-photon
		entanglement distribution via a telecom network},} arXiv 1710.09150  (2017).

\bibitem{Maring2017}
N.~Maring, P.~Farrera, K.~Kutluer, M.~Mazzera, G.~Heinze, and H.~de~Riedmatten,
\enquote{{Photonic quantum state transfer between a cold atomic gas and a
		crystal},} Nature \textbf{551}, 485--488 (2017).

\bibitem{Pelc2011a}
J.~S. Pelc, L.~Ma, C.~R. Phillips, Q.~Zhang, C.~Langrock, O.~Slattery, X.~Tang,
and M.~M. Fejer, \enquote{{Long-wavelength-pumped upconversion single-photon
		detector at 1550 nm: performance and noise analysis},} Optics Express
\textbf{19}, 21445 (2011).

\bibitem{Ates2012}
S.~Ates, I.~Agha, A.~Gulinatti, I.~Rech, M.~T. Rakher, A.~Badolato, and
K.~Srinivasan, \enquote{{Two-photon interference using background-free
		quantum frequency conversion of single photons emitted by an InAs quantum
		dot},} Physical Review Letters \textbf{109}, 1--5 (2012).

\bibitem{Lenhard2017}
A.~Lenhard, J.~Brito, M.~Bock, C.~Becher, and J.~Eschner, \enquote{{Coherence
		and entanglement preservation of frequency-converted heralded single
		photons},} Optics Express \textbf{25}, 11187 (2017).

\bibitem{Fernandez-Gonzalvo2013}
X.~Fernandez-Gonzalvo, G.~Corrielli, B.~Albrecht, M.~Grimau, M.~Cristiani, and
H.~de~Riedmatten, \enquote{{Quantum frequency conversion of quantum memory
		compatible photons to telecommunication wavelengths},} Optics Express
\textbf{21}, 19473 (2013).

\bibitem{Pelc2010}
J.~S. Pelc, C.~Langrock, Q.~Zhang, and M.~M. Fejer, \enquote{{Influence of
		domain disorder on parametric noise in quasi-phase-matched quantum frequency
		converters.}} Optics letters \textbf{35}, 2804--2806 (2010).

\bibitem{Pelc2012}
J.~S. Pelc, Q.~Zhang, C.~R. Phillips, L.~Yu, Y.~Yamamoto, and M.~M. Fejer,
\enquote{{Cascaded frequency upconversion for high-speed single-photon
		detection at 1550 nm},} Optics Letters \textbf{37}, 476 (2012).

\bibitem{Afzelius2015}
M.~Afzelius, N.~Gisin, and H.~de~Riedmatten, \enquote{{Quantum memory for
		photons},} Physics Today \textbf{68}, 42--47 (2015).

\bibitem{Laplane2017}
C.~Laplane, P.~Jobez, J.~Etesse, N.~Gisin, and M.~Afzelius, \enquote{{Multimode
		and Long-Lived Quantum Correlations between Photons and Spins in a Crystal},}
Physical Review Letters \textbf{118}, 1--5 (2017).

\bibitem{Kutluer2017}
K.~Kutluer, M.~Mazzera, and H.~de~Riedmatten, \enquote{{Solid-State Source of
		Nonclassical Photon Pairs with Embedded Multimode Quantum Memory},} Physical
Review Letters \textbf{118}, 210502 (2017).

\bibitem{Hensen2015}
B.~Hensen, H.~Bernien, A.~E. Dr{\'{e}}au, A.~Reiserer, N.~Kalb, M.~S. Blok,
J.~Ruitenberg, R.~F.~L. Vermeulen, R.~N. Schouten, C.~Abell{\'{a}}n,
W.~Amaya, V.~Pruneri, M.~W. Mitchell, M.~Markham, D.~J. Twitchen, D.~Elkouss,
S.~Wehner, T.~H. Taminiau, and R.~Hanson, \enquote{{Loophole-free Bell
		inequality violation using electron spins separated by 1.3 kilometres},}
Nature \textbf{526}, 682--686 (2015).

\bibitem{Rutz2017}
H.~R{\"{u}}tz, K.~H. Luo, H.~Suche, and C.~Silberhorn, \enquote{{Quantum
		Frequency Conversion between Infrared and Ultraviolet},} Physical Review
Applied \textbf{7}, 1--7 (2017).

\bibitem{Allgaier2016}
M.~Allgaier, V.~Ansari, L.~Sansoni, V.~Quiring, R.~Ricken, G.~Harder,
B.~Brecht, and C.~Silberhorn, \enquote{{Highly efficient frequency conversion
		with bandwidth compression of quantum light},} Nature Communications
\textbf{8}, 1--6 (2016).

\bibitem{Ikuta2014}
R.~Ikuta, T.~Kobayashi, S.~Yasui, S.~Miki, T.~Yamashita, H.~Terai, M.~Fujiwara,
T.~Yamamoto, M.~Koashi, M.~Sasaki, Z.~Wang, and N.~Imoto, \enquote{{Frequency
		down-conversion of 637 nm light to the telecommunication band for
		non-classical light emitted from NV centers in diamond},} Optics Express
\textbf{22}, 11205 (2014).

\bibitem{Ikuta2011}
R.~Ikuta, Y.~Kusaka, T.~Kitano, H.~Kato, T.~Yamamoto, M.~Koashi, and N.~Imoto, \enquote{{Wide-band quantum interface for visible-to-telecommunication wavelength conversion.}} Nature communications \textbf{2}, 1 (2011).

\bibitem{Rutz2016a}
H.~R{\"{u}}tz, K.~H. Luo, H.~Suche, and C.~Silberhorn, \enquote{{Towards a
		quantum interface between telecommunication and UV wavelengths: design and
		classical performance},} Applied Physics B: Lasers and Optics \textbf{122},
1--8 (2016).

\bibitem{Roussev2004}
R.~V. Roussev, C.~Langrock, J.~R. Kurz, and M.~M. Fejer, \enquote{{Periodically
		poled lithium niobate waveguide sum-frequency generator for efficient
		single-photon detection at communication wavelengths},} Optics Letters
\textbf{29}, 1518 (2004).

\bibitem{Gundogan2015}
M.~G{\"{u}}ndo{\u{g}}an, P.~M. Ledingham, K.~Kutluer, M.~Mazzera, and
H.~de~Riedmatten, \enquote{{Solid State Spin-Wave Quantum Memory for Time-Bin
		Qubits},} Physical Review Letters \textbf{114}, 230501 (2015).

\bibitem{Rielander2014}
D.~Riel{\"{a}}nder, K.~Kutluer, P.~M. Ledingham, M.~G{\"{u}}ndo{\u{g}}an, J.~Fekete,
M.~Mazzera, and H.~de~Riedmatten, \enquote{{Quantum storage of heralded
		single photons in a praseodymium-doped crystal},} Physical Review Letters
\textbf{112}, 1--5 (2014).

\bibitem{Seri2017}
A.~Seri, A.~Lenhard, D.~Riel{\"{a}}nder, M.~G{\"{u}}ndo{\u{g}}an, P.~M. Ledingham,
M.~Mazzera, and H.~de~Riedmatten, \enquote{{Quantum Correlations between
		Single Telecom Photons and a Multimode On-Demand Solid-State Quantum
		Memory},} Physical Review X \textbf{7}, 021028 (2017).

\bibitem{Zaske2011}
S.~Zaske, A.~Lenhard, and C.~Becher, \enquote{{Efficient frequency
		downconversion at the single photon level from the red spectral range to the
		telecommunications C-band.}} Optics express \textbf{19}, 12825--36 (2011).

\bibitem{Rielander2016}
D.~Riel{\"{a}}nder, A.~Lenhard, M.~Mazzera, and H.~de~Riedmatten,
\enquote{{Cavity enhanced telecom heralded single photons for spin-wave solid
		state quantum memories},} New Journal of Physics \textbf{18}, 123013 (2016).

\bibitem{Fekete2013}
J.~Fekete, D.~Riel{\"{a}}nder, M.~Cristiani, and H.~de~Riedmatten,
\enquote{{Ultranarrow-band photon-pair source compatible with solid state
		quantum memories and telecommunication networks},} Physical Review Letters
\textbf{110}, 1--5 (2013).

\bibitem{Kobayashi2016}
T.~Kobayashi, R.~Ikuta, S.~Yasui, S.~Miki, T.~Yamashita, H.~Terai, T.~Yamamoto,
M.~Koashi, and N.~Imoto, \enquote{{Frequency-domain Hong-Ou-Mandel
		interference},} Nature Photonics \textbf{10}, 441--444 (2016).

\bibitem{Fasel2004}
S.~Fasel, O.~Alibart, S.~Tanzilli, P.~Baldi, A.~Beveratos, N.~Gisin, and
H.~Zbinden, \enquote{{High-quality asynchronous heralded single-photon source
		at telecom wavelength},} New Journal of Physics \textbf{6}, 1--11 (2004).

\bibitem{Raymer2010}
M.~Raymer, S.~van~Enk, C.~McKinstrie, H.~McGuinness, \enquote{{Interference of two photons of different color},} Optics Communications \textbf{283}, 747-752 (2010).

\bibitem{Clemmen2016}
S.~Clemmen, A.~Farsi, S.~Ramelow, A.~Gaeta, \enquote{{Ramsey Interference with Single Photons},} Physical Review Letters \textbf{117}, 1--6 (2016).

\bibitem{Dreau2018}
A.~Dr{\'{e}}au, A.~Tcheborateva, A.~Mahdaoui, C.~Bonato, R.~Hanson, \enquote{{Quantum frequency conversion to telecom of single photons from a nitrogen-vacancy center in diamond},} arXiv 1801.03304 (2018).

\end{thebibliography}

\end{document}